\documentclass{aastex631}

\usepackage{booktabs}

\usepackage[version=4]{mhchem} 

\shorttitle{AASTeX v6.3.1 Sample article}
\shortauthors{Wang et al.}
\graphicspath{{./}{figures/}}

\begin{document}

\title{Extremely energetic EUV late phase of a pair of C-class flares caused by a non-eruptive sigmoid}

\correspondingauthor{Ya Wang}
\email{wangya@pmo.ac.cn}
\author[0000-0003-3699-4986]{Ya Wang}
\affiliation{Key Laboratory of Dark Matter and Space Astronomy, Purple Mountain Observatory, CAS, \\
Nanjing, 210023, People's Republic of China \\
}

\affiliation{SUPA School of Physics \& Astronomy, University of Glasgow, G12 8QQ,UK \\
}

\author{Sargam M. Mulay}
\affiliation{SUPA School of Physics \& Astronomy, University of Glasgow, G12 8QQ,UK \\
}

\author{Lyndsay Fletcher}
\affiliation{SUPA School of Physics \& Astronomy, University of Glasgow, G12 8QQ,UK \\
}
\affiliation{Rosseland Centre for Solar Physics, University of Oslo, P.O.Box 1029 Blindern, No-0315 Oslo, Norway \\
}

\begin{abstract}

The EUV late phase is the second increase of the irradiance of the warm coronal lines during solar flares, and has a crucial impact on the Earth's ionosphere. In this paper, we report on the extremely energetic EUV late phase of a pair of C-class flares (SOL2012-06-17T17:26:11) observed on 2012 June 17 in NOAA active region 11504 
by the \textit{Atmospheric Imaging Assembly} (AIA) instrument on board the \textit{Solar Dynamics Observatory} (SDO). 
The light curves integrated over the flaring region 
show that a factor of 4.2 more energy is released in the ``warm'' (2$-$3$\times 10^6$~K) temperature passbands (e.g. AIA 335 \AA)  during the late phase than during the main peaks. The origin of the emission in this extremely energetic EUV late phase is a non-eruptive sigmoid situated in a multi-polar magnetic field configuration, which is rapidly energised by C-class flares. The sigmoid plasma appears to reach temperatures in excess of $10^7$~K, before cooling to produce the EUV late-phase emission. This is seen in high-temperature passbands (e.g. AIA 131 \AA) and by using differential emission measure analysis. Magnetic extrapolations indicate that the sigmoid is consistent with formation by magnetic reconnection between previously existing J-shaped loops.  The sigmoid experienced a fast and a slow cooling stages, both of which were dominated by conductive cooling. The estimated total cooling time of the sigmoid is shorter than the observed value. So, we proposed that the non-eruptive sigmoid, heated by the continuous magnetic reconnection, leads to the extremely energetic EUV late phase.

\end{abstract}


\section{Introduction} \label{Section1:Introduction}

Solar flares, one of the most prominent phenomena in the solar atmosphere, are widely believed to result from the rapid release of magnetic energy. The radiative response during flares enhances the electromagnetic spectrum, spanning from gamma rays to radio wavelengths. Based on the X-ray fluxes recorded by the \emph{Geostationary Operational Environmental Satellite} (GOES), flares can typically be divided into three stages, namely 1) pre-flare phase, which is observed just before the start time of the flare; 2) the impulsive phase which is defined by its hard X-ray HXR emission when available, but coincides with the rise of the GOES X-rays and 3) the gradual phase occurring after the GOES X-ray peak \citep[e.g.][]{2002A&ARv..10..313P,2009AdSpR..43..739S,2011SSRv..159...19F}.
The general picture of a flare involves the rapid release of energy during the impulsive phase, facilitated by magnetic reconnection in the corona, and resulting in heating, mass motion, and the acceleration of electrons and ions. The electrons produce intense, HXR bremsstrahlung radiation, particularly as they interact with the dense, cool plasma in the solar chromosphere and transition region. However, HXR sources are occasionally observed in the corona \citep[e.g.][]{1994Natur.371..495M,2003ApJ...595L.103K,2008A&ARv..16..155K,2018ApJ...863...83G}.
Chromospheric heating, at least in part by these high-energy particles, leads to the expansion of chromospheric plasma upwards into the corona (``evaporation''), filling the flare loops \citep{1978ApJ...220.1137A}. Consequently, radiation in soft X-ray (SXR) and extreme ultraviolet (EUV) passbands is enhanced during the gradual phase. \\

The thermal evolution of the gradual phase, after the heating and evaporation phase is over, is generally understood to be dominated by conductive and radiative cooling \citep[e.g.][]{2001SoPh..204...91A, 2002ApJ...578..590R} resulting in the fluxes in wavelength ranges sensitive to decreasing temperatures peaking sequentially as time progresses. However, \citet{2011ApJ...739...59W} showed that the gradual phase sometimes exhibits a second peak in the EUV irradiance observed at ``warm coronal'' wavelengths, but unaccompanied by peaks in higher temperature channels. This was discovered using the \textit{Extreme ultraviolet Variability Experiment} \citep[EVE;][]{2012SoPh..275..115W}, a Sun-as-a-star spectrometer onboard the \textit{Solar Dynamics Observatory} \citep[SDO;][]{Pesnell2012_SDO_book}. The authors named this phase the ``EUV Late Phase" (ELP). It refers to the emission observed in the SDO passbands dominated by 2$-$3~MK plasma, due to Fe~\textsc{xv} and Fe~\textsc{xvi} ions. The emission commonly comes from long, high coronal loops in a multipolar magnetic field \citep[][]{2013ApJ...768..150L, 2013ApJ...778..139S}. For example \citet{2023A&A...675A.147C}
identified an ELP as originating in ongoing magnetic restructuring following the eruption of a magnetic flux rope in a fan-spine topology.
\\

While the impulsive phase is usually regarded as the most important phase for energy release overall, in the context of space weather the high-energy radiation produced during both impulsive and gradual phases is significant for the Earth's ionosphere. The enhancement of EUV irradiance caused by flares has an important impact on the Earth's ionosphere and thermosphere \citep[e.g.][]{1971ApJ...164..151K,1978ApJ...222.1043D}, and sometimes the ELP produces much more EUV radiative energy than in the impulsive phase, or the intensity peak during the ELP is stronger than the impulsive phase. Therefore, the ELP should be included as a component of ionosphere-thermosphere models during flares. For example, an extremely large ELP with a ratio of $\sim$2.1 between the ELP peak and the impulsive peak, caused by the failed eruption of a hot structure, possibly the observational feature of a flux rope, was reported by \citet{2015ApJ...802...35L}. The M1.2 non-eruptive flare reported by \citet{2018ApJ...863..124D} had an ELP peak emission over 1.3 times that of the main flare peak. The peak intensity of the ELP is over 1.4 times that of the main flare peak in an X1.8 noneruptive solar flare \citep{2024ApJ...977..257L}.
 \\

The effect of the ELP  on the global total electron content (TEC), disturbing the Earth's space environment, has been investigated \citep[e.g.][]{2024ApJ...974..157L,2024ApJ...971..188B}. \citet{2024ApJ...974L..19L} studied the X9.3 class flare SOL2017-09-06 using the modified empirical Flare Irradiance Spectral Model V2.0 (FISM2), including the ELP, as input to a coupled ionosphere-thermosphere model. They found that the TEC increases during the ELP more than during the flare main phase, and that the ELP led to enhancements in electric fields and the equatorial electrojet, in agreement with observations. \\

The cause of the stronger ELP remains a mystery, specifically, whether it is related to the flare class, the magnetic field structure, or the duration of the late phase. The flares with a stronger ELP in the literature listed above are almost M- or X-class flares, and there are few studies on whether smaller flares can possess stronger ELPs. In this study, we focus on a small flare associated with a non-eruptive sigmoid, to explore the cause of the ELP and explore the thermodynamics of the ELP.\\

In many cases, flares are associated with the eruption of an S-shaped feature called a sigmoid \citep{2021MNRAS.504.1201M}, observed in soft X-rays or high-temperature EUV emission  \citep{2002ApJ...574.1021G}.
It is thought that a sigmoid is the visible signature of a flux rope or highly sheared magnetic arcade: such single structures can be formed by reconnection between a pair of pre-existing J-shaped arcades \citep[e.g.][]{2001ApJ...552..833M,2004ApJ...617..600G}. 
Sigmoids attract attention mainly because they are a useful 
precursor of a coronal mass ejection (CME; \citealt{1999GeoRL..26..627C}). However, sigmoids without an eruption tend to attract less attention than sigmoids with an eventual eruption. \\

In a previous investigation \citep{2020ApJ...905..126W}, we reported an M-class flare with EUV late-phase emission during the partial eruption of a large-scale filament. The emission was interpreted as due to additional heating from magnetic reconnection between contracting filament threads and the low-lying magnetic field. In this paper, we report an extremely energetic EUV late phase in C-class flares associated with a non-eruptive sigmoid. In Section~\ref{section2:data_instruments}, we describe the data, instruments, and methods, while the observation and analysis are presented in Section~\ref{section3:results}. Discussion and conclusions are given in Section~\ref{Section4:conclusion}.\\

\section{The data, instruments, and methods}
\label{section2:data_instruments}

We obtained the Soft X-ray data from GOES and HXR data from the \textit{Reuven Ramaty High Energy Solar Spectroscopic Imager} \citep[RHESSI;][]{Lin2002} using the relevant SolarSoft (SSW) GUI interfaces \citep{Freeland1998SoPh..182..497F}. The isothermal temperature during a solar flare is calculated using the flux ratio of 0.5$-$4 \AA\ to 1$-$8 \AA\ channels, based on the method described in \citet{2005SoPh..227..231W}. The RHESSI data is not available before 17:20~UT or after 18:10~UT due to satellite night and the South Atlantic Anomaly (SAA), respectively. 
\textcolor{black}{
We reconstruct RHESSI images
using the CLEAN algorithm \citep{Hurford2002SoPh..210...61H} in energy bands at 3$-$6 and 6$-$12~keV during the late-phase time interval of 17:58$-$18:03~UT. 
The CLEAN algorithm starts with a dirty map created by back projection and iterates towards an optimal model-data agreement based on the assumption that the image can be well represented by a superposition of point sources, convolved with the RHESSI beam (point-spread function).\footnote{\url{https://hesperia.gsfc.nasa.gov/rhessi3/software/imaging-software/clean/index.html}}}\\
\begin{figure*}
\begin{center}
\includegraphics[trim=0cm 0cm 0.5cm 0.5cm, width=0.95\textwidth]{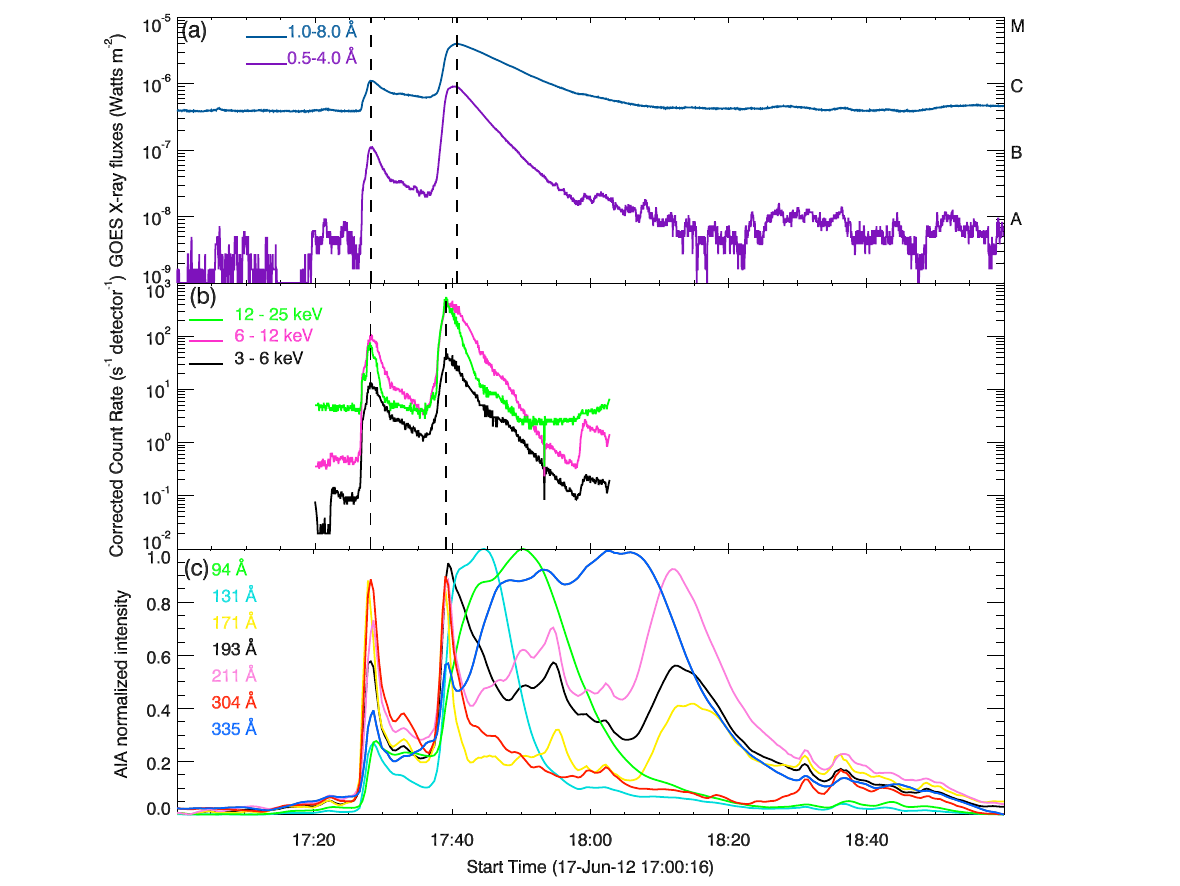}
\caption{Panel (a): The SXR fluxes observed by GOES in 1$-$8 (blue) and 0.5$-$4.0 \AA\ (purple) passbands for the C1.0 and C3.9 flares on 2012 June 17.  Panel (b): The X-ray fluxes (corrected count rates) from RHESSI in various energy ranges from 3 to 25~keV. Panel (c): The normalised intensity in 7 EUV passbands of AIA observed during these two C-class flares. The normalized intensity was calculated for the flaring region within the large white box in panels a1 and b1 of Fig.~\ref{fig:fig2}. The vertical dashed lines indicate the respective times for the peak fluxes observed by GOES (panel a) and RHESSI (12$-$25 keV, panel b) during these flares.}

\label{fig1:GOES_rhessi_aia}
\end{center}
\end{figure*}

The \textit{Atmospheric Imaging
Assembly} \citep[AIA;][]{2012SoPh..275...17L} instrument onboard SDO takes spatially resolved observations in ten passbands with a pixel size of 0$\farcs$6 and a cadence of 12~s and 24~s \citep{2012SoPh..275...17L}. The ten passbands have different temperature response functions \citep{ODwyer2010, DelZanna2011, DelZanna2013}. 
(Note, ``temperature'' here refers to the electron temperature, though it is often assumed that all species have the same temperature.)
Of these, six EUV passbands are sensitive to coronal temperatures: 94\,\AA\ (with main contribution from Fe~\textsc{xviii}, $\log T$\,[K] $\sim$ 6.8), 131\,\AA\ (Fe~\textsc{viii, xxi}, $\log T$\,[K] $\sim$ 5.6, 7.0), 171\,\AA\ (Fe~\textsc{ix}, $\log T$\,[K] $\sim$ 5.8), 193\;\AA\ (Fe~\textsc{xii, xxiv}, $\log T$\,[K]  $\sim$ 6.2, 7.3), 211\;\AA\ (Fe~\textsc{xiv}, $\log T$\,[K]  $\sim$ 6.3), and 335\;\AA\ (Mg~\textsc{viii}, $\log T$\,[K] $\sim$5.9,  Fe~\textsc{xvi}, $\log T$\,[K] $\sim$6.4). 
During flares, 193\;\AA\ may in fact be dominated by Fe~\textsc{xxiv} at $\log T$\,[K] $\sim$ 7.3, and there are expected to be contributions at the $\sim$ 10\% level from Ca~\textsc{xvii} and Ca~\textsc{xvi} (both $\log T$\,[K] $\sim$ 6.8) in the 193\,\AA\ and 211\,\AA\ channels respectively. 
High-temperature continuum is also important:
free-free continuum from 10-20\,MK plasma in the 211\,\AA\ and 171\,\AA\ channels and helium free-bound continuum in the 211\,\AA\ channel may also contribute at the level of tens of percent, though this is apparently less significant for smaller flares \citep{2013ApJ...777...12M}. The 304\;\AA\ passband (He~\textsc{ii}, $\log T$\,[K] $\sim$4.7) is sensitive to chromosphere and transition region temperatures. The flare excess in the 1600\,\AA\ and 1700\,\AA\ passbands is dominated by chromospheric and transition region lines formed at temperatures in the range $4.2 < \log T [K] < 5.1$ \citep{2019ApJ...870..114S}. \textcolor{black}{We downloaded full-disk AIA level 1 data from the  \textit{Joint Science Operations Center}(JSOC\footnote{\url{http://jsoc.stanford.edu/ajax/lookdata.html}}) and converted this to level 1.5 by using the \texttt{aia\_prep.pro} routine in SSW. For these flare events, SDO/EVE data is not available. } \\

\textcolor{black}{
To study the spatial distribution and evolution of the thermal structure of the sigmoid, we performed a differential emission measure (DEM) analysis using the SSW `\texttt{xrt\_dem\_iterative2.pro}' method \citep{2004IAUS..223..321W}. To remove the influence of data saturation and diffraction fringes during the impulsive phase of the two C-class flares and the EUV late phase, we desaturated the six AIA EUV channels by using the Sparsity-Enhancing DESAT (SE-DESAT) method \citep{2019ApJ...882..109G} available in SSW. 
We calculated the AIA temperature response function for the EUV channels employing the method described by \citet{2011A&A...535A..46D}, using CHIANTI version v10.1  \citep{Del_Zanna2021ApJ...909...38D} and an electron number density of 1$\times$ $10^{10}~\mathrm{cm}^{-3}$. \\} 

\textcolor{black}{To study the magnetic structure of the active region, we use a series of data sets of magnetic field provided by the \textit{Helioseismic and Magnetic Imager} (HMI, \citealt{2012SoPh..275..229S}) instrument onboard SDO. }
\textcolor{black}{We carried out \textit{Non-Linear Force-Free Field} (NLFFF) extrapolation, adopting the optimisation approach proposed by \cite{2000ApJ...540.1150W} and implemented by \cite{2004SoPh..219...87W} and \cite{2006SoPh..233..215W}. The hmi.B$_{-}$720s data series, which is HMI full-disk disambiguated vector fields, is used in the NLFFF extrapolation. The 180-degree ambiguity in the transverse components of the vector magnetic field is resolved by the improved minimum energy method\footnote{The reader can refer to \url{https://www.cora.nwra.com/AMBIG/}} \citep [e.g.,][]{2006SoPh..237..267M, 2009SoPh..260...83L}. The information in the `disambig' segment of the data files indicates whether the azimuth should be increased by 180 degree. To combine the disambig and azimuth information, we use `hmi\_disambig.pro' in SSWIDL.} \\\begin{figure*}
\begin{center}
\includegraphics[trim=0cm 0.7cm 0cm 1.2cm, width=0.95\textwidth]{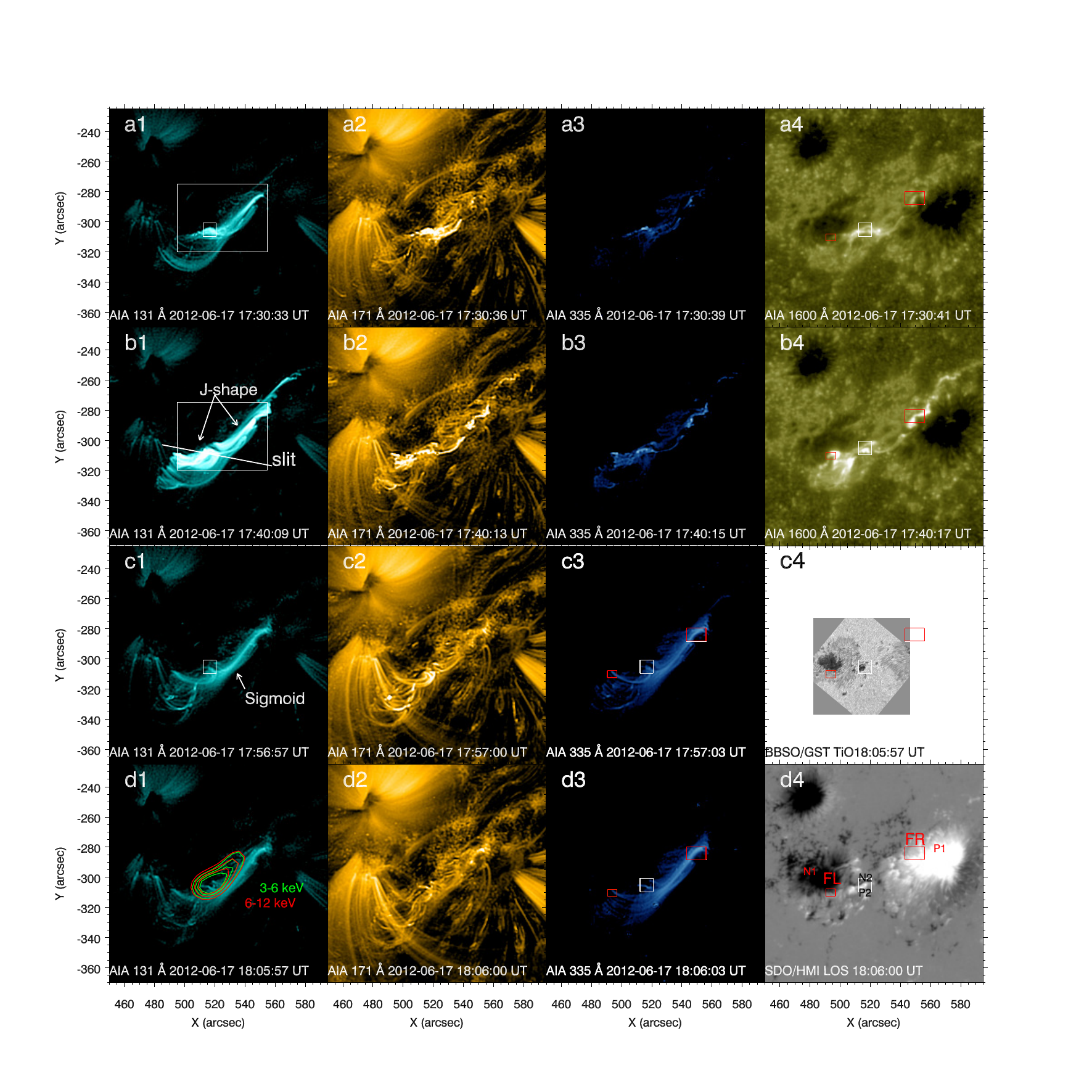}
\caption{The evolution of EUV emission observed in three EUV passbands of AIA at 131, 171, and 335~\AA~during the peak time of the C1.0 (panels a1-a3) and C3.9 (panels b1-b3) flares. The images in panels c1-c3 and d1-d3 were taken during the EUV late phase. The UV counterparts of these two flares were observed in the AIA 1600~{\AA} passband (panels a4 and b4). 
The region outlined by the large white box in panels a1 and b1 was used to obtain total intensity in the flaring location, with the resulting profiles shown in panel (c) of Fig.~\ref{fig1:GOES_rhessi_aia}. Small white boxes in panels a1, c1, a4, and b4 indicate the location where the magnetic reconnection might have occurred, and the red boxes in panels a4, b4, c3-c4, and d3-d4 show the footpoint of the sigmoidal structure that is indicated by FL (left footpoint) and FR (right footpoint). A white diagonal line in panel b1 indicates the position of an artificial slit which was used to create the space-time diagrams in Fig.~\ref{fig:fig3}. Panel c4: The field-of-view observed by the TiO broadband filter in 7057\,{\AA}. Panel d1: The green and red contours represent the X-ray sources at 3$-$6~keV and 6$-$12~keV energy channels of RHESSI. Panel d4: the line-of-sight (LOS) magnetogram observed by HMI. The two footpoints of the sigmoid are rooted in the two sunspots, the leading (positive polarity, P1) and trailing (negative polarity, N1). Between these two sunspots, we observe a pore at the location of a dipolar field labelled as N2 and P2. An animation of the evolution of the flare is available. The animation starts at June 17th, 2012 at 17:20 UT. It ends the same day around 19:45 UT. The real-time duration is 42 seconds. In the animation top row from top left to right is the SDO/HMI LOS, SDO/AIA 94\AA, SDO/AIA 131\AA, SDO/AIA 171\AA. The bottom rows is SDO/AIA 1600\AA, SDO/AIA 335\AA, SDO/AIA 304\AA, and a graph of the GOES lightcurves in 1$-$8 \AA\ (blue) and 0.5$-$4.0 \AA\ (pink) for the C1.0 and C3.9 flares on 2012 June 17. The vertical line indicates the timeline of each snapshot.}
\label{fig:fig2}
\end{center}
\end{figure*}

High-resolution ground-based images in He~\textsc{i} 10830 \AA\ and TiO passbands are obtained from the 1.6-meter aperture  \textit{Goode Solar Telescope} (GST; \citealt{Goode2010, 2012SPIE.8444E..03G}) at the  \textit{Big Bear Solar Observatory} (BBSO). They are used to study the chromospheric and photospheric response during the flares.
The He~\textsc{i} 10830~\AA\  filter is a narrowband filter with a bandpass of 0.5~\AA~and captures images every 10~s, with a pixel size of 0$\farcs$0875 and a field of view (FOV) of 90$\arcsec\times$90$\arcsec$. The TiO 7057~\AA\ filter is a broadband filter with a bandpass of 10~\AA \ and a pixel size of 0$\farcs$034.
The co-alignment of the BBSO/GST data, as well as between BBSO/GST and SDO, is performed using cross-correlation. Using the Fast Fourier Transform, we calculated the cross-correlation coefficient between two images, which were first rebinned to the same resolution, at various relative shifts.\\

\section{Observation and Analysis} \label{section3:results}
\subsection{Overview of the flaring active region} \label{subsection3.1:overview}

\textcolor{black}{These flares occurred in NOAA active region 11504 (S17~W41) on 2012 June 17. The Solar Object Locator (SOL; \citealt{Leibacher2010}) for this event is SOL2012-06-17T17:26:11\footnote{\url{https://www.lmsal.com/hek/her?cmd=view-voevent&ivorn=ivo://helio-informatics.org/FL_FlareDetective-TriggerModule_20120617_174411_2012-06-17T17:26:11.070_1}}. GOES-15 recorded C1.0 and C3.9 class flares in the soft X-ray (SXR) 1$-$8 and 0.5$-$4~{\AA} passbands. The GOES X-ray fluxes for these flares are shown in panel (a) of Fig.~\ref{fig1:GOES_rhessi_aia}. The C1.0 flare started at 17:26~UT and peaked at 17:28~UT. The C3.9 flare started at 17:36~UT and peaked at 17:41~UT. There is a precursor at $\sim$17:23~UT before the start of the first C1.0 flare}. The peak GOES times for both flares are indicated by the black dashed lines in panel (a) of Fig.~\ref{fig1:GOES_rhessi_aia}
\textcolor{black}{
and time profiles in energy ranges 3$-$6, 6$-$12 and 12$-$25~keV from RHESSI are shown in panel (b). They show two peaks at  17:28 and 17:39:02~UT in the energy bands 6$-$12 and 12$-$25~keV. The first peak in RHESSI coincides with the GOES-15 SXR peak time of the C1.0 flare, but the second peak, shown by the black dashed line in panel (b) of Fig.~\ref{fig1:GOES_rhessi_aia}, appeared one and a half minutes before the GOES peak time of the C3.9 flare, consistent with the Neupert effect \citep{Neupert68, Hudson91, Dennis93}.}\\

\textcolor{black}{The AIA light curves shown in Fig.~\ref{fig1:GOES_rhessi_aia}(c) are obtained by integrating intensities over the sigmoid region within the large white boxes shown in panels a1 and b1 of Fig.~\ref{fig:fig2}. }
We studied the evolution of the flaring region in 7 (2) EUV (UV) channels, and a few snapshots during two C-class flares are shown in panels a1-a3, b1-b3, c1-c3, and d1-d3 (a4 and b4) of Fig.~\ref{fig:fig2}. The EUV images at 131\,\AA\ showed bright J-shaped loops prior to the second flare that evolve into a sigmoidal loop structure seen in panels c1 and c3. The morphology of this bright emission suggests an underlying J-shaped magnetic field.
We identified the ends of the sigmoidal loops in AIA 335 and 1600\,{\AA} images and highlighted them as small red boxes in panels c3, d3, a4, and b4. We overplotted these boxed regions on the TiO~7057\,\AA\ photospheric image and the HMI line-of-sight (LOS) magnetogram images to identify their photospheric counterparts. These are shown in panels c4 and d4. These images revealed that the left footpoint (FL) and right footpoint (FR) of the sigmoidal loops were embedded in the two sunspots with opposite magnetic polarity; the leading sunspot has a positive magnetic field (P1), whereas the trailing one has a negative magnetic field (N1). In addition, we observed a pore with a dipolar field between these two sunspots (labelled as N2 and P2 in panel d4). This magnetic structure, characterised by a parasitic polarity within the large-scale bipolar field, underwent rapid evolution. The brightenings and plasma outflows that are seen between the two sets of J-shaped loops are commonly regarded as the observational evidence of magnetic reconnection.\\

The CDAW CME catalogue\footnote{\url{https://cdaw.gsfc.nasa.gov/CME\_list/}} \citep{Gopalswamy_LASCO_catalog2024}  of the \textit{Large Angle and Spectrometric Coronagraph} (LASCO) on board the \textit{Solar and Heliospheric Observatory satellite} (SoHO) \citep{Brueckner1995_LASCO} indicated that the flares were accompanied by a weak CME.  
However, the H$\alpha$ and He~\textsc{i} 10830 \AA\ observations indicate that the CME was related to the partial eruption of a filament, which is studied in detail by \citet{2022ApJ...929...85D}. They reported that the western part of the filament remained in the J-shaped structure. While the filament material on the eastern part escaped. The sigmoid that formed moved but did not erupt. \\

\begin{figure*}
\begin{center}
\includegraphics[trim=0cm 0cm 0cm 0.5cm, width=0.95\textwidth]{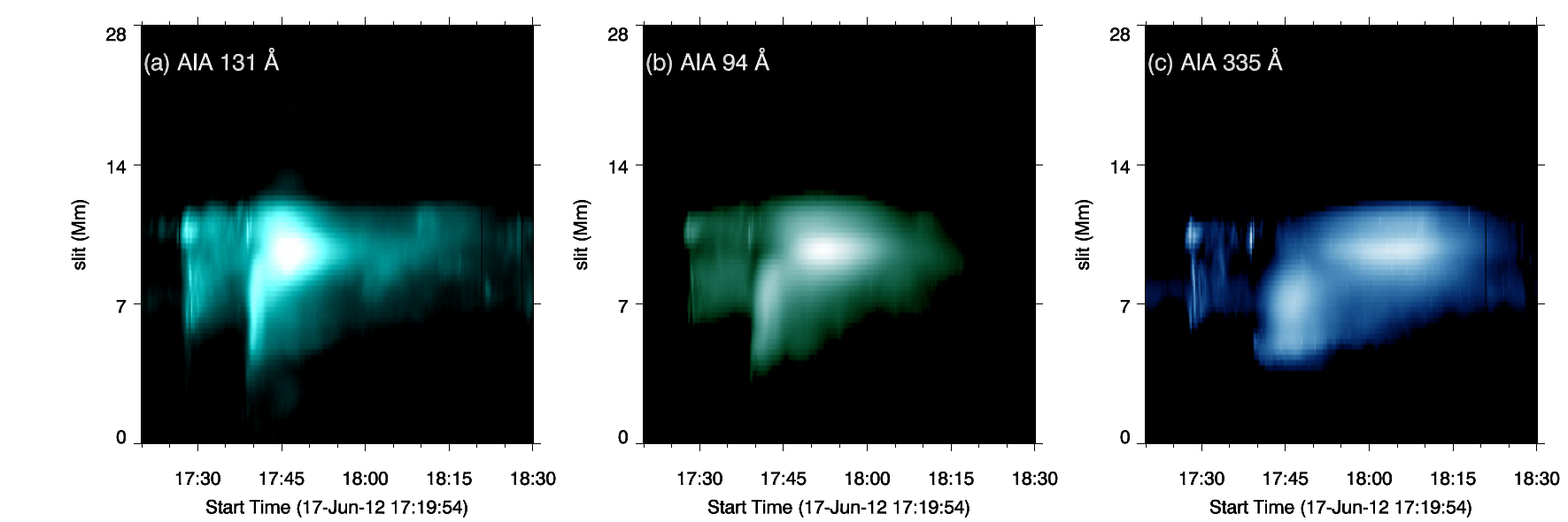}
\caption{Space-time diagrams in 94 \AA, 131 \AA\ and 335 \AA\ passbands along the slit shown as a white line in Fig.~\ref{fig:fig2} panel b1.}
\label{fig:fig3}
\end{center}
\end{figure*}

We created a time-distance diagram in Fig.~\ref{fig:fig3}, taking a cut along the white line in Fig.~\ref{fig:fig2}, panel b1, which is also the direction of motion of the sigmoid. This shows that the sigmoid stopped at about 10~Mm along this slice.
The brightest emission from the core of the sigmoid shows up consecutively at 131, 94, and 335 \AA\ images, likely indicating cooling. These channels have a double-peaked temperature response function. During the flare, 131, 94, and 335 \AA\ are dominated by Fe~\textsc{XXI} ($\log T$\,[K] $\sim$ 7.0), Fe~\textsc{XVIII} ($\log T$\,[K] $\sim$ 6.8), and Fe~\textsc{XVI} ($\log T$\,[K] $\sim$ 6.4), respectively.\\

The RHESSI source locations are overlaid on the sigmoid structure in the AIA 131~{\AA} image as green and red contours in panel d1 of Fig.~\ref {fig:fig2}.
The sources are located at the sigmoid and are consistent with the small peaks of the RHESSI lightcurves at 3$-$6 and 6$-$12~keV at around 17:59~UT, during the peak of the AIA 335 \AA\ lightcurve. \\

\subsection{Extremely energetic EUV late phase} 
\label{subsection3.1:exterme_EUV_late_phase}

\begin{figure*}
\begin{center}
\includegraphics[trim=0cm 0.5cm 0cm 0.2cm, width=0.85\textwidth]{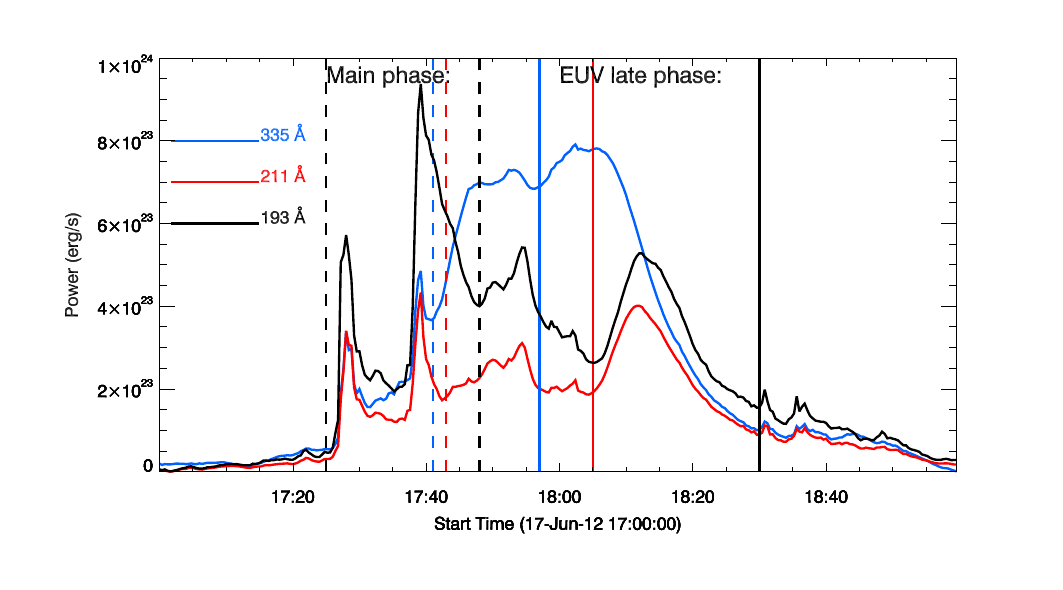}
\caption{Background subtracted   intensity profiles (in units of energy/s) observed in AIA 335 \AA\ (blue), 211 \AA\ (red), and 193 \AA\ (black)  passbands for the large white box region shown in panels a1 and b1 of Fig.~\ref{fig:fig2}. The background is obtained based on the minimum of the integrated intensity over the time period from 17:00 to 19:00 UT.
The time period of the main phase is indicated by dashed lines, starting from 17:25 UT and ending at different turning points of 17:41, 17:43, and 17:48~UT for 335, 211, and 193~{\AA}, respectively. The time period of EUV late phase is shown in solid lines, starting from 17:57~UT for 335~{\AA} and 18:05 UT for 193 and 211 \AA, respectively. Solid vertical lines (at 18:30 UT) mark the end of the EUV late phase. The total energy is obtained by integrating over each time period.
}
\label{fig:fig4}
\end{center}
\end{figure*}

The ``main phase" of the flare (see the vertical dashed lines in Fig.~\ref{fig:fig4}) covers the two strong flare peaks at 17:28~UT and 17:39~UT. As also shown in Fig.~\ref{fig1:GOES_rhessi_aia}, the first event has a rapid rise and decay in all EUV channels, while the second has a rapid rise and decay in the cooler channels of 304, 171 and 211~\AA. There is a less rapid decay after the second peak in the 193 \AA\, channel, which has a significant hot component during flares. The first flare peak in all EUV channels aligns well with the GOES peak. The second flare peak at around 17:40~UT is several seconds earlier in EUV than the GOES peak. Additional peaks appear after the second flare peak, first in the hotter passbands of 131\,\AA\ (dominated by Fe~\textsc{xxi} in flares), 94\,\AA\, and 335\,\AA, then at lower temperatures. This is consistent with plasma cooling from $\sim$17 MK, obtained from the GOES temperature, following the second flare. Starting from around 17:57~UT, a peak appears in the `warm' channel of 335\,\AA\ but is unaccompanied by increases in GOES, or channels of 94\,\AA, 131\,\AA, or 193\,\AA~which are decaying at that time. In the statistical study of \cite{2011ApJ...739...59W}, the ELP delay time after the main GOES peak ranges from 41 to 204 minutes, which is longer than the delay in this event, but the pattern of emission at different temperatures accords with two of the criteria in \cite{2011ApJ...739...59W}'s definition of an ELP,  namely, a second peak in the lightcurves of Fe~\textsc{XVI} emission after the GOES X-ray peak observed by AIA (no EVE data for this flare), and no corresponding enhancements in GOES soft X-ray or hotter emission lines.
The other two ELP criteria, of an eruptive event and longer loops forming later on, are not obeyed by this event. However, we will call this ELP emission. It is unclear whether this ELP should be associated with both flares, or only with the second, but it is likely to be related to the non-erupting sigmoid. \\

From Fig.~\ref{fig1:GOES_rhessi_aia}(c), we see that the ratio of the peak intensity observed in the 335 \AA\ passband during the EUV late phase to the intensity of the main peak of the \text{C1.0 flare is more than 2 and of the C3.9 flare is about 1.8}. This is comparable to some other flare events with an extremely large EUV late phase, for example, the ratio of 2.1 for a M1.0 class flare \citep{2015ApJ...802...35L}  and 1.3 for a M1.2 class flare \citep{2018ApJ...863..124D}.\\

To investigate the importance of EUV energy after the main flare peaks, we calculate the energy in EUV radiation using several AIA channels at 193, 211, and 335~{\AA}. The energy can be estimated following the method described by \citet{2014ApJ...793...70M}. A pre-flare background, based on the minimum of the integrated intensity over the time period from 17:00 to 19:00~UT, is subtracted from the time profile for each channel, and these background-subtracted profiles,
in units of DN s$^{-1}$, are converted into radiated power in units of erg\,s$^{-1}$, by using the expression \textit{intensity/(EA $\times$ photon energy $\times$ area)} (R.Milligan, private communication). EA represents the AIA instrument's effective area per unit wavelength for each channel, which is obtained from the AIA response functions, the \texttt{aia\_get\_response.pro} routine in SSW. 
The area is \texttt{2$\pi$$r{^2}$}, where $r$ is 1~AU in cm, assuming a uniform
angular distribution of the radiation \citep{2006JGRA..11110S14W}. 
Fig.~\ref{fig:fig4} shows the results of this. It can be seen that the time profile of the 335 \AA\ emission peaks at 7.9$\times$10$^{23}$~erg s$^{-1}$, exceeding the peak power in this passband radiated during the main flare peaks at 17:28 and 17:39~UT. 
\\

We separated the event into two phases based on the turning points of the radiated power curves, indicated by the vertical lines in Fig.~\ref{fig:fig4}. 
These turning points are different for different passbands. The main phase is defined to start at 17:25~UT (first vertical black dashed line) and ends at the turning points shown by vertical dashed lines at 17:41, 17:43, and 17:48~UT, for AIA channels 335, 211, and 193~\AA\, respectively. The EUV late phase is defined to start at the turning points marked by the vertical solid lines, at 17:57~UT for 335~\AA, and 18:05~UT for 211 and 193~\AA, respectively, and continues until 18:30~UT.
We compare the energy during the main phase (the additional peaks are not included) and the EUV late phase.
The total 335 \AA\ passband energy in the EUV late phase from 17:57 to 18:30~UT is 3.9$\times$10$^{25}$~erg, whereas during the main phase interval 17:25 to 17:41~UT it is 9.3$\times$10$^{24}$~erg. The total energy of the EUV late phase is more than four times that of the main phase in this passband, and the ratio reaches 1.8 for the 211 \AA\ passband. The energy radiated in the temperature passbands (335, 211, and 193 \AA) is summarised in Table~\ref{table1}. As reported by \citet{2024ApJ...974..157L}, the TECs in the ionosphere during the EUV late phase increased more than that during the main phase based on the results from both the observation and simulation. We speculate that the effects on the ionosphere of this event could be significant. However, further confirmation is needed in the next study, especially for such small flares in which the ELP is stronger than the main phase.

\begin{figure*}
\begin{center}
\includegraphics[trim=0cm 0cm 0cm 0cm, width=0.9\textwidth]{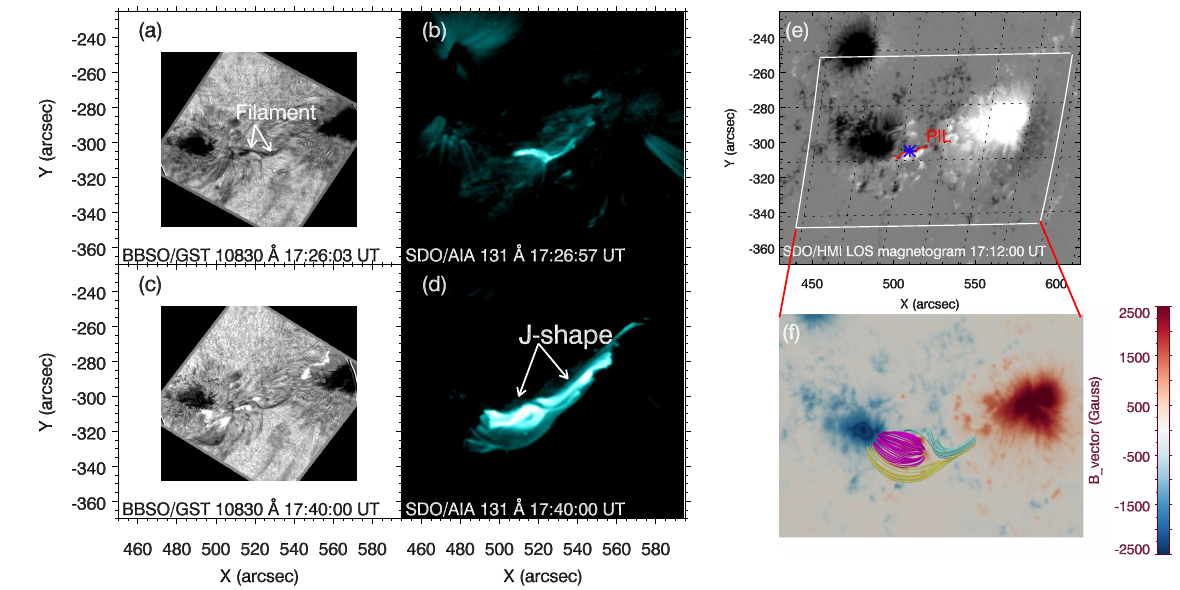}
\caption{Panels (a) and (b): images at 10830 \AA\ and 131 \AA\, representing the filament and its counterpart in high temperature at $\sim$ 17:26 UT. Panels (c) and (d) display the filament and J-shape structure observed at 10830 and 131 \AA\ respectively, at 17:40 UT. Panel (e) indicates the magnetogram along the line of sight at 17:12:00 UT. The bottom boundary is from the corresponding vector magnetogram. The white lines indicate the region of NLFFF extrapolation. The white portion represents a positive magnetic field, while the black portion represents a negative magnetic field. The red curve represents the polarity inversion line (PIL), and the blue asterisk indicates the location where we measure the decay index. Panel (f) shows magnetic topology obtained from NLFFF extrapolation seen from the Z-direction. The background shows the magnitude of the magnetic field at the plane of Z=0. The pink, cyan and yellow lines represent the magnetic field lines.}

\label{fig:fig8}
\end{center}
\end{figure*}

\begin{figure*}
\begin{center}
\includegraphics[trim=0cm 0cm 0cm 0cm, width=1.0\textwidth]{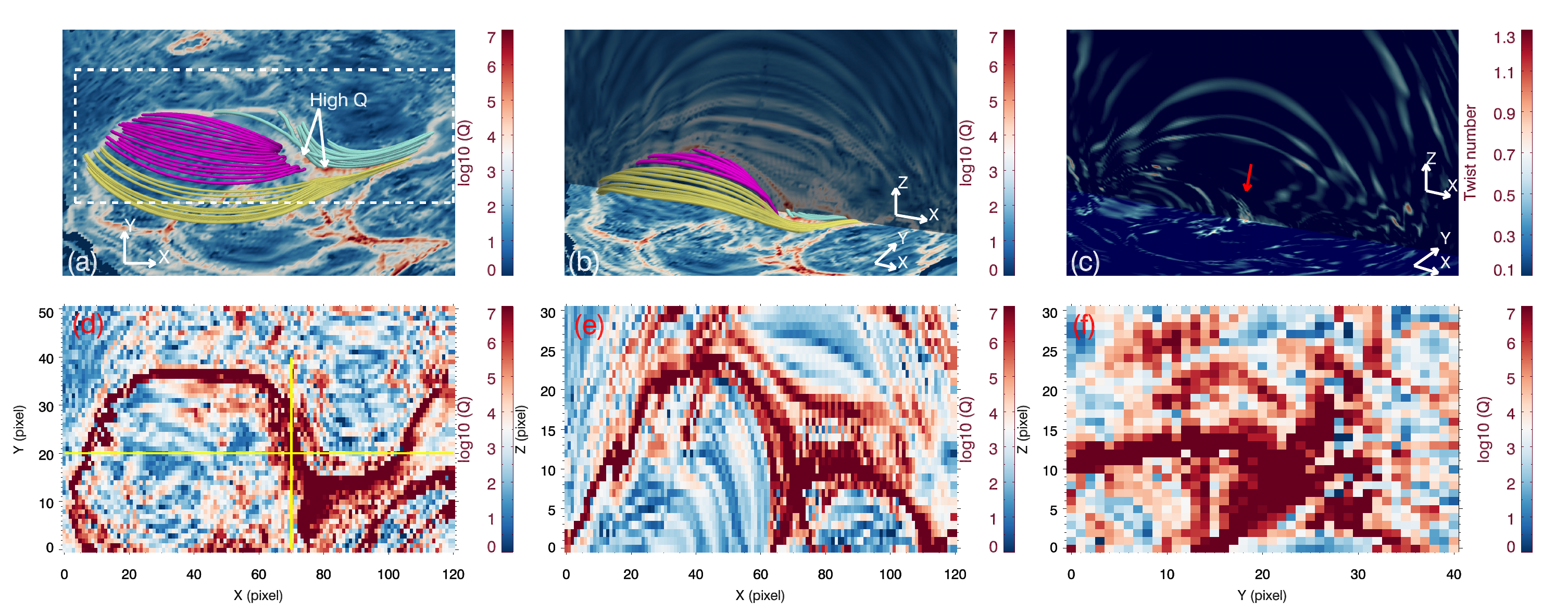}
\caption{3D magnetic topology of the J-shape and sigmoid obtained by the NLFFF extrapolation. Panel (a)-(b) Top and side views of the 3D logarithmic Q overplotted with the  J-shape and sigmoid structures. Panel (c) shows the 3D magnetic twist T from the side view. Panel (d)-(f) show the distributions of logarithmic Q in the X-Y plane based on the white box shown in panel (a), in the X-Z plane at Y-axis = 20 along the horizontal white line in panel (d), and in the Y-Z plane at X-axis = 70 along the vertical white line in panel (d), respectively.}
\label{fig:fig9}
\end{center}
\end{figure*}
\begin{figure*}
\begin{center}
\includegraphics[trim=0cm 0cm 0cm 0cm, width=0.6\textwidth]{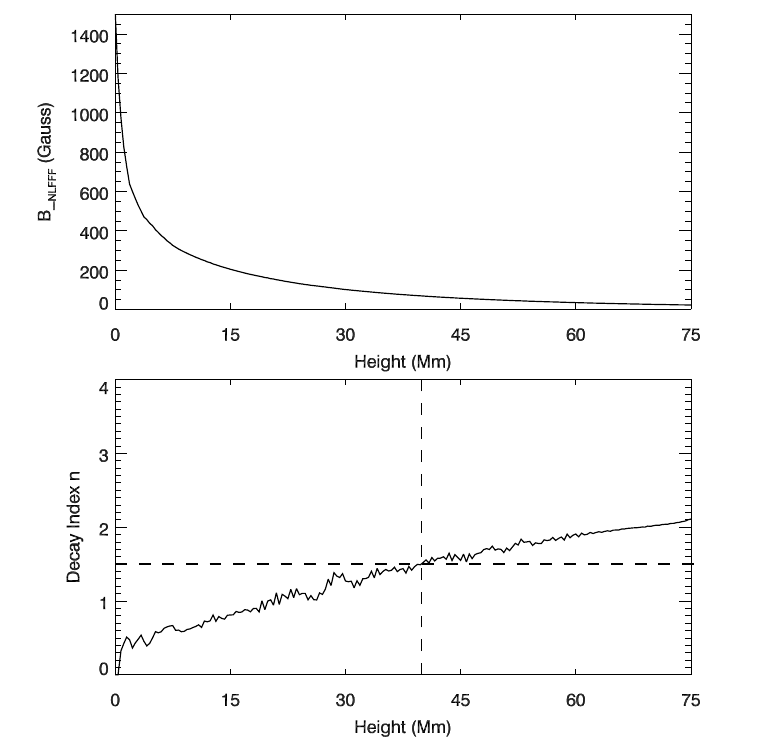}
\caption{Top panel: magnitude of the horizontal magnetic field above the location indicated by a blue asterisk in Fig.~\ref{fig:fig8} (panel e); Bottom panel: the corresponding decay index with height from the NLFFF model. When the decay index equals 1.5, the height is about 40 Mm, indicated by the horizontal and vertical dashed lines.}
\label{fig:fig11}
\end{center}
\end{figure*}

\begin{figure*}
\begin{center}
\includegraphics[trim=0cm 0cm 0cm 0cm, width=0.95\textwidth]{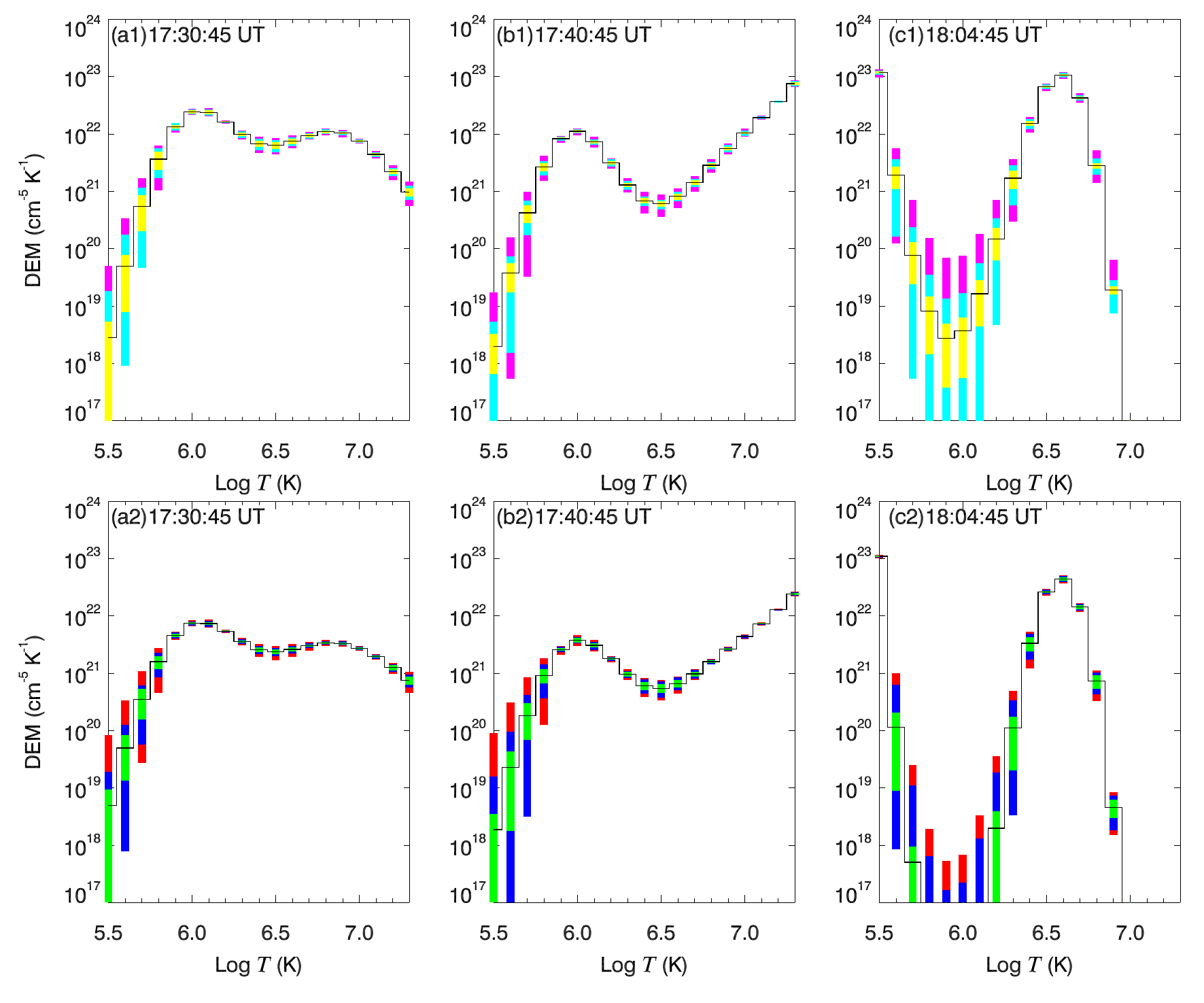}
\caption{The DEM distribution profiles for the center of the sigmoid, which is shown as a white box in Fig.~\ref{fig:fig5} panel c3. In the top (bottom) panel, the DEMs are obtained using the photospheric (coronal) elemental abundances.
Panel a1-a2: DEMs at 17:30:45~UT (C1.0 flare), Panel b1-b2: 17:40:45~UT (C3.9 flare), and Panel c1-c2: 18:04:45~UT (EUV late phase). The black profiles display the best-fitted DEM curves. The colored bars (yellow, turquoise, and pink in the top row, green, blue and red in the bottom row) represent the 50\%, 80\%, and 95\% uncertainties associated with DEMs obtained from 100 Monte Carlo solutions. 
}
\label{fig:fig62}
\end{center}
\end{figure*}

\begin{figure*}
\begin{center}
\includegraphics[trim=0cm 0.7cm 0cm 1.2cm, width=0.95\textwidth]{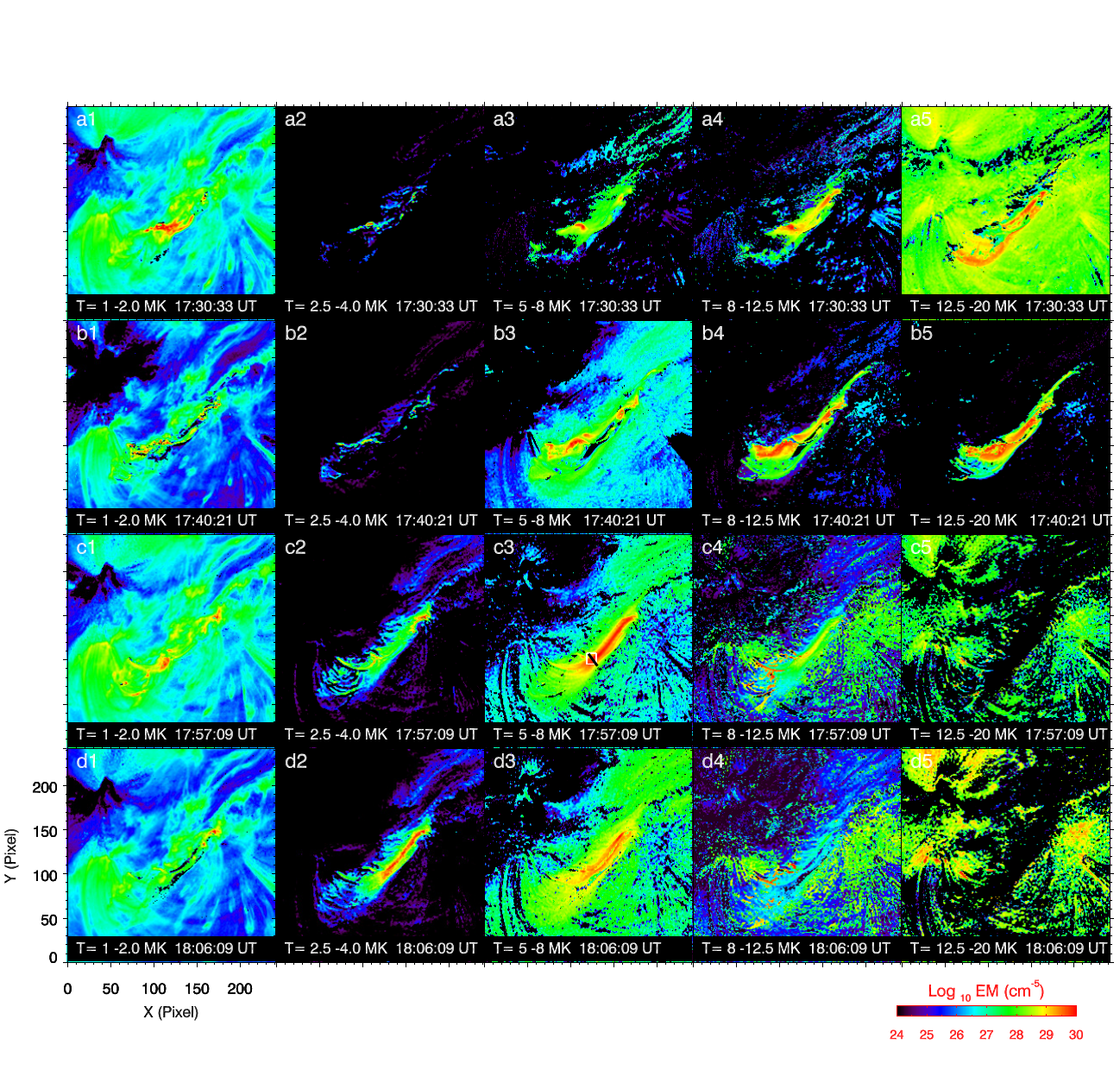}
\caption{Emission measure maps derived from DEM analysis by using the photospheric element abundances. The vertical columns, from left to right, show the EM integrated over four different temperature bands: 1$-$2.5 MK, 2.5$-$4.5 MK, 5$-$8 MK, 8$-$12.5 MK, and 12.5$-$20 MK. The rows from top to bottom correspond to four different times: 17:30:33~UT, 17:40:21~UT, 17:57:09~UT, and 18:06:09~UT, respectively. The top of the sigmoid region highlighted in the white box in panel (c3) was used to obtain the DEMs in Fig.~\ref{fig:fig62}. The short black line in panel (c3) indicates the position where the width of the sigmoid obtained. }
\label{fig:fig5}
\end{center}
\end{figure*}
\begin{figure*}
\begin{center}
\includegraphics[trim=0cm 0cm 0cm 0cm, width=0.75\textwidth]{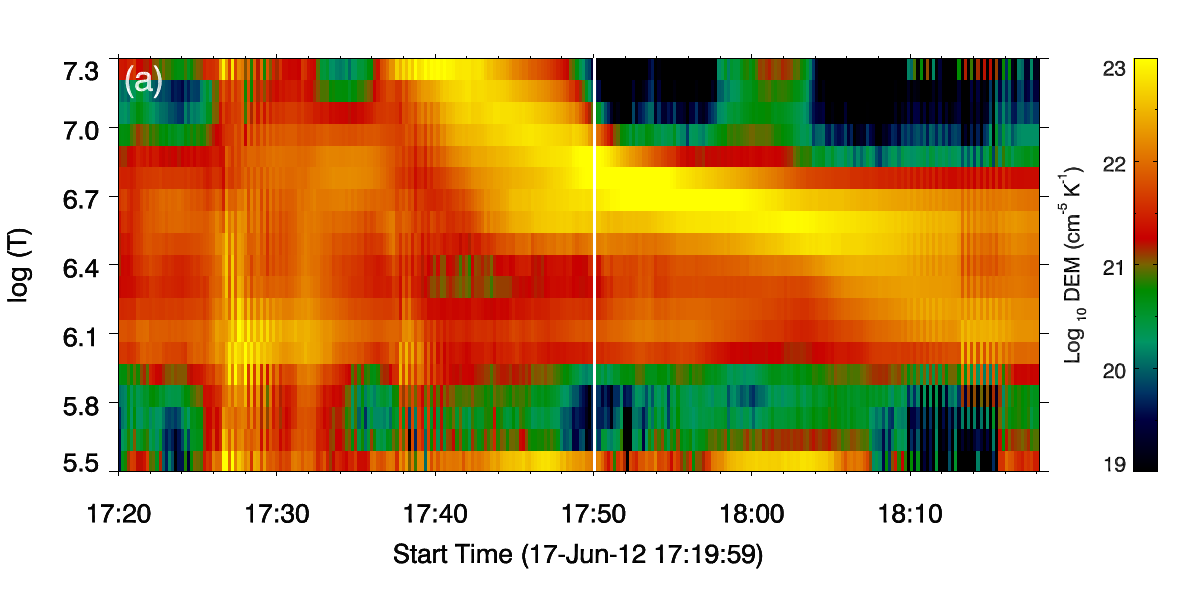}
\caption{The emission measure distribution as a function of time and temperature, represented by the horizontal X-axis and vertical Y-axis, respectively. The DEM (using the photospheric element abundances) is obtained for the small white box shown in Fig.~\ref{fig:fig5}c. The white vertical line at 17:50 UT indicates the turning point of the two stages of the cooling process.}
\label{fig:fig61}
\end{center}
\end{figure*}

\begin{table}[htbp]
\centering
\caption{Energy radiated in the AIA 335 \AA, 211 \AA\ and 193 \AA\ passbands. 
}
\setlength{\tabcolsep}{0.9mm}{ 
\begin{tabular}{c|c|c|c}
\hline 
Passbands &335 \AA\ &211 \AA\ &193 \AA\  \\
\hline 

Flare stage& \multicolumn{1}{|c|}{main phase}  late phase & \multicolumn{1}{|c|}{main phase}late phase  &  \multicolumn{1}{|c|}{main phase}{late phase}  \\
\hline

Energy (erg) & \multicolumn{1}{|c|} {$9.3\times 10^{24} $}$3.9\times 10^{25} $  &  \multicolumn{1}{|c|} {$8.2\times 10^{24} $} $1.5\times 10^{25}$ &  \multicolumn{1}{|c|}{$2.4\times 10^{25} $} $2.1\times 10^{25}$   \\
\hline
\end{tabular}\label{table1}}
\end{table}

\subsection{Magnetic topology and formation of the sigmoid}
\label{subsection3.4topology}

The magnetic topology is key to understanding the role of magnetic reconnection in a solar flare, as well as the subsequent evolution of flare loops and ribbons. In a typically low $\beta$ plasma, such as in the corona, where the ratio of gas pressure to magnetic pressure is low, we investigate the magnetic topology of the flare-associated active region via NLFFF extrapolation. \\

The region for the extrapolation is indicated in Fig.~\ref{fig:fig8} panel (e), and the adopted dimensions are 124$\times$73$\times$73 Mm$^{3}$. The bottom boundary is the photospheric vector magnetic field at 17:12:00~UT obtained from the SDO/HMI before the occurrence of the C class flares. The resulting magnetic topology of this targeted region is shown in Fig.~\ref{fig:fig8} panel (f). By referring to the location of the sigmoid at 131 \AA\ and sunspots at 1600 \AA, we repeatedly selected pairs of points that correspond to the location of the sigmoid endpoints. The extrapolated magnetic field is calculated using the 4th-order Runge-Kutta integrator in the ParaView software. The pink magnetic arcade corresponds to the eastern J-shaped arcade seen at 131~\AA, although the extrapolated field does not have a clear J-shape. The cyan field corresponds to the western J-shaped arcade. The overlying magnetic loop system shown in yellow, constraining the sigmoid structure, could explain its the non-eruptive nature. To strengthen this explanation, we measure the decay index, defined as $n=-dlog(B_{h})/dlog(h)$, from the NLFFF model. We selected a location (shown as a blue asterisk) along the PIL in Fig.~\ref{fig:fig8} panel (e), where we obtained the horizontal magnetic field and the decay index using the same method described in \citet{2018ApJ...859..148W} and the results are shown in Fig.~\ref{fig:fig11}. Figure 6(e) shows the low-lying nature of the sigmoid, and we estimated the height to be about 11 Mm (with a resolution of 0.5 \arcsec/pixel). The decay index is less than 1 below 15 Mm. The critical index for the torus instability ranges from 1$-$2 in theoretical calculations and numerical simulation \citep{2006PhRvL..96y5002K}. Furthermore, for a value of the decay index equal to 1.5, which is the value for a toroidal current channel \citep{1978mit..book.....B}, the height of around 40 Mm is beyond the sigmoid. Thus, the sigmoid does not reach the height of the threshold of the torus instability, and it is non-eruptive.\\

As described in section 3.1, the TiO 7057 \AA\ image and the magnetogram indicate a multipolar magnetic field consisting of two sunspots with opposite magnetic fields (P1 and N1) and a pore with a dipolar field (P2 and N2) between them. By comparing the images at 10830 and 131 \AA\ shown in Fig.~\ref{fig:fig8}(c,d) with the extrapolation, we see that the western J-shaped arcade is co-spatial with 
a filament in He~\textsc{i} 10830 \AA\ passband. For the footpoints of the pink and cyan arcades, representing the two J-shaped arcades visible in 131\AA, one footpoint of each is located in one of the sunspots (the positive spot P1 for the (cyan) western J-shape arcade and the negative spot N1 for the (pink) eastern arcade) and the other is located in the pore (with negative pore field N2 for the western J-shape arcade and positive pore field P2 for the eastern arcade). The configuration of the magnetic field in the extrapolation is consistent with magnetic reconnection occurring between the magnetic arcade shown in pink and the J-shaped field in cyan.  \\

To further examine the possibility of magnetic reconnection between these two sets of loops, we calculate the squashing factor $\textit{Q}$ of the reconstructed magnetic field by using the method described in \citet{2016ApJ...818..148L}, and obtain the 3D logarithmic $\textit{Q}$ distribution maps in the photosphere and vertical planes along X- and Y-axes shown in Fig.~\ref{fig:fig9}. The squashing factor $\textit{Q}$ is defined by \citet[]{2002JGRA..107.1164T} and quasi-separatrix layers (QSLs) are quantified by high $\textit{Q}$ values. Since the QSLs often refer to magnetic topological discontinuities where strong electric fields are produced, and in turn current sheets, they are one of the favourable locations for magnetic reconnection \citep{2014masu.book.....P}. Additionally, \citet{1996A&A...308..643D} conjectured that QSLs are the regions where the electric current density increases, and thus could possibly be the location for the occurrence of magnetic reconnection. As can be seen in the lower panels of Fig.~\ref{fig:fig9}, the $\textit{Q}$ factor between the two sets of loops is above 3, which is higher than the surroundings, indicating the possibility of magnetic reconnection in the region where the $\textit{Q}$ factor is large. Consequently, the sigmoid could be formed via magnetic reconnection between the two J-shape structures seen in 131\,\AA\ images.\\ 

To detect the twist of the sigmoid, we calculate the twist number $\textit{T}_{\omega}$ in the region by using the method described in \citet[]{2016ApJ...818..148L}. The twist number can be used to measure how many turns two infinitesimally close field lines wind about each other \citep[e.g.,][]{2006JPhA...39.8321B,2016ApJ...818..148L}. As shown in panel (c) of Fig.~\ref{fig:fig9}, the maximum of the twist number of the sigmoid is around 0.8. The relatively low twist number of the sigmoid, less than 1$-$2 turns, may be expected for a newly-forming sigmoid structure via continuous magnetic reconnection between two J-shaped structures.\\

\subsection{Thermal structure of the sigmoid} \label{sec-dem}

Fig.~\ref{fig:fig62} shows the DEMs calculated at a location at the center (where the two J-shape structures connected) of the sigmoid, indicated by the white box shown in Fig.~\ref{fig:fig5} panel c3, chosen
as representative of the sigmoid characteristics in each stage of the flare. The black solid line shows the best-fit DEM curve consistent with the observed fluxes. The uncertainties on the DEM solutions were obtained from 100 Monte Carlo realisations of the data, using 3$\%$ of the observational value as the Gaussian sigma.
The three timings to be analysed are around the impulsive peaks of the two C-class flares (17:30:45 and 17:40:45~UT) and the peak of the EUV late phase (18:04:45~UT) observed in 335~\AA. Fig.~\ref{fig:fig62} panels (a1)-(c1)  show the DEM distribution over the temperature range $5.5 < \log T\,\mathrm{[K]} < 7.3$ ($\sim$0.3 MK to 20 MK) calculated using photospheric element abundances, with panels (a2)-(c2) showing the results for coronal abundances. These indicate the multi-thermal nature of plasma at the flaring sigmoid location. These DEMs were measured to compare our peak values of DEMs and the corresponding peak temperatures obtained using two elemental abundances. Regardless of the choice of abundances, the DEM distribution profiles show similar shapes and peak temperatures (in panels a1 and a2; panels b1 and b2; panels c1 and c2), but the DEM values differ by a factor of 2-4. Differences are likely due to the varying relative abundances of Ca and Fe, resulting in distinct contributions from the Ca lines.
\citet{2014ApJ...786L...2W} investigated the elemental abundance in 21 solar flares and found that the flare composition is close to the photospheric, suggesting that the bulk of the plasma evaporated during a flare comes from deep below the chromosphere. So, we proceed to analyse the EM map (Fig.~\ref{fig:fig5}) and the DEM distribution (Fig.~\ref{fig:fig61}) as a function of time and temperature using photospheric elemental abundances.
\\

For the C1.0 flare, at 17:30:45~UT the DEM showed two peaks, one at $\log T\,[\mathrm{K}] = 6$ (1~MK) and a second at $\log T\,[\mathrm{K}] =6.8$ (6.3~MK), whereas the DEM for the C3.9 flare at 17:40:45~UT showed a peak at $\log T\,[\mathrm{K}] = 6$  and an apparent increase of the DEM to temperatures over $10^7$\,K. During the EUV late phase at 18:04:45~UT  
the DEM peaked at high-temperature $\log T\,[\mathrm{K}] = 6.6$ (4~MK). 
The temperature interval used to run the DEMs shown in this plot is from  $\log T\,\mathrm{[K]}$ = 5.5--7.3. However, the shape of the DEM distribution and the peak DEMs and corresponding peak temperatures do not vary even though we vary the upper limit ($\log T\,\mathrm{[K]}$ = 7.1--7.5) for the temperature used in the DEM analysis. \\

Note that the DEMs are not well constrained in the low ($\log T\,[\mathrm{K}] < 5.7$) and high-temperature ranges ($\log T\,[\mathrm{K}] > 7$), meaning that the increase towards high temperatures seen at 17:40:45~UT must be viewed with caution. However, additional evidence for the presence of plasma at $\log T\,[\mathrm{K}] > 7$ comes from a comparison of the 171\,\AA, 193\,\AA\ and 211\,\AA\ light curves in Fig.~\ref{fig1:GOES_rhessi_aia} between 17:40~UT and 17:50~UT. In non-flaring conditions these are dominated by Fe lines formed at $\log T\,[\mathrm{K}]\sim$ 5.8, 6.2 and 6.3, respectively, but the 193\,\AA\ channel also includes a contribution from Fe~XXIV at $\log T\,[\mathrm{K}]\sim 7.3$. Starting from 17.40~UT, the 171\,\AA\ and 211\,\AA\ light curves have a similar rapid decrease over 5 minutes or less, whereas the 193\,\AA\ light curve falls more slowly, over the same 10-minute interval where the DEM suggests cooling from $\log T\,[\mathrm{K}]\sim 7.3$. This may be an indication that during this time the Fe~XXIV contribution to the 193\,\AA\ channel is significant. The behaviour of the emission measure distribution seen between 17:40~UT and 17:50~UT in Fig.~\ref{fig:fig61} is also consistent with cooling from an initially high temperature. In addition, we compared the observed intensities in six AIA channels with the predicted intensities obtained from the best-fit DEM solution to check the reliability of the DEM calculation in terms of the validity of the hot component of the sigmoid. The comparison is shown in Table 2, and the ratio in the AIA 193 \AA\ channel is 1.025-1.150 at 17:30:45 and 17:40:45 UT, a difference which is not very significant.
\\

To study the spatial distribution and evolution of the sigmoid in different temperature ranges, we follow the approach of \citet{2021MNRAS.504.1201M} who analysed the temperature structure and evolution of sigmoidal regions. We calculate the emission measure (EM) by using the photospheric element abundances over five different temperature ranges using 
$\mathrm{EM}=\int \mathrm{DEM}(T) dT$
and obtain emission measure maps by calculating this for each pixel in the region of interest. Fig.~\ref{fig:fig5} shows the emission measure maps at four different times, and in temperature bands, 1$-$2.0~MK, 2.5$-$4.0~MK, 5$-$8~MK, 8$-$12.5~MK and 12.5$-$20~MK. 
At 17:40:21 UT, the sigmoid is dominated by emission at temperatures in the 12.5$-$20\,MK range.
At 17:57:09~UT the dominant emission is in the 5$-$8\,MK range, reducing to 2.5$-$4.0~MK by 18:06:09~UT. \\

Fig.~\ref{fig:fig61} shows the DEM distribution over time and temperature obtained for the averaged intensities in the white box shown in Fig.~\ref{fig:fig5} panel c3. The enhancements at around 17:30~UT and 17:40~UT correspond to the two C-class flares. The first flare shows two peaks at $\log T\,[\mathrm{K}] \sim 6$ and $\log T\,[\mathrm{K}] \sim 6.8$, and the second shows the DEM starting  above $\log T\,[\mathrm{K}] \sim 7$, decreasing to $\log T\,[\mathrm{K}] \sim 6.7$ at 17:50~UT. 
During the EUV late phase from 17:50~UT to 18:10~UT, the differential emission measure is dominated by a temperature range of 2.5 to 5~MK.\\

\begin{table}[htbp]
\centering
\caption{Observed and predicted intensities from DEM (Unit: DN $s^{-1}$) and the ratio of observed and predicted intensities}
\begin{tabular}{cccccc}
\hline
     17:30:45 UT &Obs. &DEM predict (photospheric) &Ratio&DEM predict (coronal)&Ratio  \\
     \hline
     AIA 131&181.9&182.4 & 0.997&182.8 & 0.995\\
     AIA 94&48.9&49.4 & 0.989&49.2 & 0.994\\
     AIA 211&590.8&609.7 & 0.969&608.0 & 0.972\\
     AIA 193&2037.7&1983.3 & 1.027&1987.9 & 1.025\\
     AIA 171&2070.2&2082.1 & 0.994&2078.8 & 0.996\\
     AIA 335&34.1&31.1 &1.096 &31.4 & 1.086\\
     \hline
     17:40:45 UT &Obs. &DEM predict (photospheric) &Ratio&DEM predict (coronal)&Ratio\\
     \hline
     AIA 131&536.5&602.4 & 0.891&582.7 & 0.921\\
     AIA 94&51.9&43.6 & 1.190&47.1 & 1.102\\
     AIA 211&260.0&275.4 & 0.944&268.2 & 0.969\\
     AIA 193&2893.9&2517.1 & 1.150&2660.4 &1.088 \\
     AIA 171&1082.9&1078.4 & 1.004&1081.7 & 1.001\\
     AIA 335&25.4&25.0 & 1.016&25.38 & 1.001\\     
     \hline
     18:40:45 UT &Obs. &DEM predict (photospheric) &Ratio&DEM predict (coronal)&Ratio\\
     \hline
     AIA 131&102.3&126.2 & 0.811&107.7 & 0.950\\
     AIA 94&57.6&55.4 &1.040 &59.2 & 0.973\\
     AIA 211&435.3&468.9 & 0.928&446.8 & 0.974\\
     AIA 193&1411.7&906.3 &1.558 &924.9 & 1.526\\
     AIA 171&417.9&414.7 & 1.008&458.8 & 0.911\\
     AIA 335&110.1&96.8 & 1.137&111.6 &0.987 \\ 
     \hline

\end{tabular}

\end{table}

\subsection{Cooling time estimation}

To investigate the cooling process of the sigmoid, we divided the cooling process into two phases from 17:40 to 17:50~UT and from 17:50 to 18:10~UT, between which there is an obvious turning point at around 17:50 UT as seen from Fig.~\ref{fig:fig61}. The rapid cooling of the sigmoid from 17:40 to 17:50~UT can be seen from the time profiles of SDO/AIA EUV flux variations (Fig.~\ref{fig1:GOES_rhessi_aia} (c)), in which the peaks of the light curves during the rapid cooling process appear sequentially from 131 \AA, 94 \AA, 335 \AA, 211~\AA\ to 171 \AA, over an interval of around 15~minutes. We can estimate the cooling time from the DEM evolution, by using the equation $T(t)=T_{0}\exp(-t/\tau_\mathrm{cool})$, where $T_{0}$ is the initial temperature at time $t=0$, and $\tau_{cool}$ is the cooling time. Based on the DEMs shown in Figs.~\ref{fig:fig62} and~\ref{fig:fig61}, the initial temperature at 17:40~UT is set as $\log T [K] = 7.2$ (T = 15.8\,MK) and at $t = 600 ~\mathrm{s}$ later (17:50~UT) it is $\log T [K] = 6.8$ (T = 6.3\,MK). So $\tau_\mathrm{cool}\sim 650\,\mathrm{s}$. For the second phase from 17:50 to 18:10~UT, the temperature from $\log T [K] = 6.8$ (T = 6.3\,MK) at 17:50~UT cools down $\log T [K] = 6.2$ (T = 1.6\,MK) at 18:10~UT, So $\tau_\mathrm{cool}\sim 875\,\mathrm{s}$ at the second phase.\\

The initial fast cooling of hot sigmoid loops is likely to be by thermal conduction. The conductive cooling timescale, $\tau_\mathrm{cond}$, is given approximately by \citep{2004ApJ...605..911C}:
\begin{equation}
    \tau_\mathrm{cond} = \frac{3n k_\mathrm{B}L^2}{\kappa_0 T_0^{5/2}},
\end{equation}
and the radiative cooling timescale, $\tau_\mathrm{rad}$, is given by
\begin{equation}
    \tau_\mathrm{rad} = \frac{3 k_\mathrm{B}T_0^{3/2}}{\chi n},
\end{equation}

where $n$ is the loop number density, $L$ its half-length, $T_0$ its initial isothermal temperature, and $k_\mathrm{B}$ and $\kappa_0$ are respectively the Boltzmann constant 
and the classical thermal (Spitzer) conductivity coefficient ({9.2$\times$$10^{-7}\mathrm{erg\,s^{-1}\,cm^{-1}\,K^{-7/2}}$). The loop's projected (full) length $2L$ is around 70\arcsec or $\sim$ 50,000\,km. Taking into account its solar location, its deprojected full length is $\sim$ 70,000\,km, giving a lower limit to $L_{9}$ of 3.5. 
$\chi$ is set as $10^{-18.81}$ based on the temperature (T $>$ $10^{5.11}$), which is determined by the radiative loss function \citep{1978ApJ...220..643R}.
The density can be estimated from $\sqrt{\mathrm{{EM}/{w}}}$, where $\mathrm{w}$ represents the width of the loop, indicated by the short black line overlaid on the sigmoid (Fig.~\ref{fig:fig5} (c3)).\\

At $t$=17:40~UT, the EM value in the small box equals 3.7$\times$$10^{29}$$cm^{-5}$, and the width of the sigmoid is about 2.5~Mm. So the plasma density is about 3.85$\times$$10^{10}cm^{-3}$, and the weighted-average temperature is about $\log T [K] = 7.2$ (T = 15.8\,MK). Based on the equations (1) and (2), we obtained that $\tau_\mathrm{cond}=214$ s and $\tau_\mathrm{rad}=573$ s. Thus, at the initial cooling phase, the conductive cooling timescale is shorter than the radiative cooling timescale, and the sigmoid is assumed to cool purely by conduction from the initial temperature of $\log T [K] = 7.2$ (T = 15.8\,MK). At $t = 17:50$~UT, the EM value equals to 4.6$\times$$10^{29}$$cm^{-5}$,  and the width of the sigmoid is about 3 Mm. So, the plasma density is about 3.92$\times$$10^{10}cm^{-3}$. The weighted-average temperature is about $\log T [K] = 6.8$ (T = 6.3\,MK), and the weighted-average temperature at 18:10 UT is about $\log T [K] = 6.2$ (T = 1.6~MK). We also estimated the conductive and radiative cooling timescales of the sigmoid at 17:50~UT, which are 2169 s and 2317 s, respectively. The conductive cooling still dominates the second cooling phase, but the conductive cooling timescale is much closer to the radiative cooling timescale compared to the first cooling phase. The parameters of EMs, sigmoid widths, and the weighted-average temperatures are listed in Table~\ref{DEM_parameters}. \\

The total cooling time can be estimated by \citep{2013ApJ...778...68R}:
\begin{equation}
    t_\mathrm{tot} = \tau_\mathrm{c0} \bigg[\bigg(\frac{\tau_{r0}}{\tau_{c0}}\bigg)^{7/12}-1\bigg]+\frac{2\tau_{r0}}{3}\bigg(\frac{\tau_{c0}}{\tau_{r0}}\bigg)^{5/12} \times\bigg[1-\bigg(\frac{\tau_{c0}}{\tau_{r0}}\bigg)^{1/6}\bigg(\frac{T_L}{T_0}\bigg)\bigg],
\end{equation}

Where the subscript `0' refers to the values of that property at the start of the cooling phase and ${T_L}$ indicates the final temperature at the end of the cooling phase. $\tau_r$ and $\tau_c$ imply radiative and conductive cooling timescales. As a result, the conductive cooling dominates the two cooling stages. The conductive cooling timescale is shorter than the empirical cooling timescale at the first stage, while that at the second cooling stage is longer than the empirical value. The estimated total cooling time is about 409~s, which is much shorter than the observed cooling time. Thus, we suggest that the non-eruptive sigmoid, heated by ongoing magnetic reconnection at the second cooling stage, leads to an extremely energetic EUV late phase.\\

\begin{table}[htbp]
\centering
\caption{The parameters of the EM, sigmoid width, the plasma density, the weighted-average temperature and the timescales}
\begin{tabular}{cccccccc}
\hline
      &EM ($ 10^{29}$$cm^{-5} K^{-1}$) &widths (Mm) & n $(10^{10} cm^{-3})$&Temperature (MK) & $\tau_{cond} ~(s)$ & $\tau_{rad}~ (s)$\\
     \hline
     17:40 UT& 3.7  & 2.5&3.85& 15.8  & 214  & 573\\
     17:50 UT& 4.6  & 3& 3.92 & 6.3 &2169&2317\\
     18:10 UT& /& /&/&1.6&/&/\\
     \hline

\end{tabular}\label{DEM_parameters}

\end{table}

\section{Discussion and Conclusions} \label{Section4:conclusion}

We have analysed two consecutive C-class flares that occurred on June 17, 2012, and had an extremely energetic EUV late phase. The late phase can be identified as the second peak 
at 335~\AA\, based on the SDO/AIA lightcurve integrated over the flaring area. Combining the AIA data, including a DEM analysis, with He~\textsc{i} 10830\,\AA\ observations, and results from an NLFFF extrapolation, our main findings can be summarised as follows: \\

\begin{enumerate}
    \item{Based on the light curves of SDO/AIA in the 335\,\AA\ passband, we find a second peak from 17:57 to 18:30~UT after the flare main phase, which we identify as an ELP. 
The intensity ratio of the peaks in the EUV late phase and the main peaks of the flare is $\sim$1.8 for the C3.5 flare and $\sim$2 for the C1.0 flare. Based on the radiated energy distribution, the during the EUV late phase is four times higher than that during the flare main phase in warm-temperature passband 335 \AA (During flares, the other ``warm'' channels of 193 \AA\ and  211 \AA\ have contributions from Fe~\textsc{xxiv} at $\sim$20 MK, and continuum emission at 10$-$20 MK, respectively.
The event is therefore an extremely energetic ELP.} 
\item{
After the second flare a sigmoid forms with a temperature above 10~MK. This sigmoid then cools down rapidly to 5$-$8 MK at 17:57 UT and then to 2.5$-$4 MK at 18:06 UT. The emission measure from the sigmoid between $\log T [\mathrm{K}] \sim 6.4-6.8$ dominates the EUV late phase. By investigating the cooling timescale and the total cooling time of the sigmoid, we compared the conductive and radiative cooling timescales to the empirical timescale, and found a rapid cooling at the first stage and a slow cooling process at the second stage. Besides, we compared the conductive and radiative cooling timescale and found that the conductive cooling timescale is shorter than the radiative cooling timescale in both cooling stages, indicating that the conductive cooling dominates the cooling process. By estimating the total cooling time, we found it is shorter than the observed cooling time, indicating a continuous heating in the cooling stage. }    \item{The host region is situated in a multipolar magnetic field. It presents a large-scale bipolar field with a parasitic dipole field embedded in the middle. During the two flares, two J-shaped structures are visible in 131 \AA\ images with temperatures reaching as high as 10~MK. They are co-spatial with two filaments observed in He~\textsc{i} 10830~\AA. Based on the multiple-wavelength observation and the magnetic structure, the flare region reveals a filament-hosting sigmoid in an asymmetric quadrupole magnetic field. The source of the ELP emission is a sigmoid that is consistent with being formed by continuous magnetic reconnection between two sets of J-shaped loops, which is favoured by the magnetic configuration and undergoes non-eruption.}
\item{A non-linear force-free magnetic field extrapolation shows two magnetic arcades, corresponding to the two sets of J-shaped loops. One set is rooted in the negative field of the large-scale bipolar field, and the other rooted in the positive field of the large-scale bipolar field, with the other ends in the parasitic dipole field.
These are in a region close to the pore where the $\textit{Q}$ factor is large and the twist number of the sigmoid is around 0.8. The topology would allow the sigmoid to be formed through magnetic reconnection between these sets of loops.}
\end{enumerate}

This study shows that an extremely energetic EUV late phase can be formed in C-class flares. During the ELP in this event, the peak EUV intensity is around twice that of the impulsive phase in the AIA `warm' channels, and the EUV energy radiated in the ELP is comparable to that in the flare main phase. This can be contrasted with our previous study of a M1.5 class flare with a weak EUV late phase reported in \citet{2020ApJ...905..126W}, in which a hot channel, observed in high-temperature EUV passbands, and potentially a magnetic flux rope \citep{2012NatCo...3..747Z}, erupted successfully. In that event, part of the filament erupted, and some of the filament material fell back to the lower atmosphere along with the contraction of the magnetic field lines. The additional heating caused by the interaction between the contracting field line and low-lying magnetic arcades contributed to the EUV late phase. The ratio of peaks in the late and impulsive phases in that event was only 0.56. \\

In short, a partial eruption of the filament occurs in both flare events, but the extremely energetic EUV late phase only forms in the event with the non-eruptive high-temperature magnetic sigmoid in this paper. This study is in accordance with the statistical result that the relative peaks of the non-eruptive flares are stronger than those of eruptive flares \citep{2016ApJS..223....4W}. Additionally, \citet{2018ApJ...863..124D} reported a case study on an extremely energetic EUV late phase in a non-eruptive solar flare. It will be interesting to investigate further the relationship between the strong EUV late phase and non-eruptive flares. A statistical investigation of additional non-eruptive flare events is required to establish whether energetic EUV late phases, sustained by continuous magnetic reconnection, are common in non-eruptive solar flares.  
\\

\begin{nolinenumbers}
\begin{acknowledgments}
\nolinenumbers   
\modulolinenumbers[0]
\resetlinenumber[1]
We thank our referee whose careful reading and constructive suggestions have improved this paper. This work is supported by the Strategic Priority Research Program of the Chinese Academy of Sciences, grant No. XDB 0560000 (XDB0560102), the National Key R\&D Program of China 2021YFA1600502 (2021YFA1600500), and NSFC grants 12273101, 12403068, 12173092, 12273115, and 12073081. The work is also supported by the China Scholarship Council and the Youth Fund of JiangSu No. BK20241707. SMM and LF acknowledge support from grants ST/T000422/1 and ST/X000990/1 made by UK Research and Innovation's Science and Technology Facilities Council (UKRI-STFC). We thank the SDO/AIA and SDO/HMI teams for providing the valuable data. The AIA and HMI data are downloaded via the Joint Science Operations Center (JSOC). We acknowledge the use of the RHESSI Mission Archive available at https://hesperia.gsfc.nasa.gov/rhessi/mission-archive. We gratefully acknowledge our use of
data from the Goode Solar Telescope (GST) of the Big Bear Solar Observatory (BBSO). BBSO operation is supported by the US NSF AGS-2309939 and AGS-1821294 grants and the New Jersey Institute of Technology. GST operation is partly supported by the Korea Astronomy and Space Science Institute and Seoul National University. This CME catalog is generated and maintained at the CDAW Data Center by NASA and The Catholic University of America in cooperation with the Naval Research Laboratory. SOHO is a project of international cooperation between ESA and NASA.
\end{acknowledgments}
\end{nolinenumbers}


\end{document}